\documentclass[preprint,aps,floatfix,showpacs,nofootinbib,endfloats]{revtex4}
\usepackage{amsmath}
\usepackage{graphicx}
\def\prn#1{{\left(#1\right)}}

\begin{document}
\title{Unusually large polarizabilities and ``new" atomic states in Ba}
\author{Chih-Hao Li}
\email{chihhao@socrates.berkeley.edu}
\author{S. M. Rochester}
\email{simonkeys@yahoo.com}
\affiliation{Department of Physics, University of California at
Berkeley, Berkeley, California 94720-7300}
\author{M. G. Kozlov}
\email{mgk@MF1309.spb.edu} \affiliation{Petersburg Nuclear Physics
Institute, Gatchina, 188300, Russia}
\author{D. Budker}
\email{budker@socrates.berkeley.edu} \affiliation{Department of
Physics, University of California at Berkeley, Berkeley,
California 94720-7300} \affiliation{Nuclear Science Division,
Lawrence Berkeley National Laboratory, Berkeley, California 94720}
\date{\today}
\begin{abstract}
Electric polarizabilities of four low-J even-parity states and
three low-J odd-parity states of atomic barium ranging from
$35,600\ $ to $36,000\ $cm$^{-1}$ are investigated. The states of
interest are excited (in an atomic beam) via an intermediate
odd-parity state with a sequence of two laser pulses. The
odd-parity states can be excited due to the Stark-induced mixing
with even-parity states. The polarizabilities are measured via
direct spectroscopy on the second-stage transition. Several states
have tensor and scalar polarizabilities that exceed the values
that might be expected from the known energy levels of barium by
more than two orders of magnitude. Two of the Stark-induced
transitions cannot be identified from the known energy spectrum of
barium. The observations suggest the existence of as yet
unidentified odd-parity energy states, whose energies and angular
momenta are determined in the present experiment. A tentative
identification of these states as [Xe]6s8p$\ ^3$P$_{0,2}$ is
suggested.
\end{abstract}

\maketitle

\section{Introduction}
\label{Section:intro}

Polarizabilities describe how the energy levels of a system, such
as an atom, shift in an external electric field. They provide a
way to probe the atomic wave function at large distances from the
nucleus, and are related to various physical quantities, including
van der Waals constants, dielectric constants, indices of
refraction, and charge-exchange cross sections. In this work, we
measured the polarizabilities of seven energy levels of barium in
between $35,600$ and $36,000\ $cm$^{-1}$. Although
polarizabilities of several levels of barium have been measured
\cite{ Rin97,Li95,Kul84,Kul83,van83}, those of many other states
are not yet determined. Our primary motivation for this work is to
obtain information that will be used in the analysis of
sensitivity and possible systematic uncertainties in experiments
searching for violation of Bose-Einstein statistics for photons
\cite{Eng00}.

In the first part of this work, we measured the scalar and tensor
polarizabilities and, in some cases, hyperpolarizabilities of four
even-parity states, 6p$^2\ ^3$P$_2$, 6s7d$\ ^3$D$_{1,2}$, and
5d6d$\ ^3$D$_1$, of barium (Fig.~\ref{fig:SmallRangeEd}). We found
that three out of these four states have unusually large
polarizabilities that could only be explained by the existence of
as-yet-unidentified closely-lying states of odd parity.

In the second part of the work, we looked for Stark-induced
transitions to the odd-parity states ranging from $35,600$ to
$36,000\ $cm$\ ^{-1}$. Several Stark-induced resonances were found
and classified into three sets, corresponding to transitions to
electric-field-split Zeeman sublevels of three odd-parity states.
One of these can be identified as the 6s8p$\ ^3$P$_1$ state. The
other two cannot be identified with the known energy levels of
barium and are possibly the 6s8p$\ ^3$P$_{0,2}$ states, which are
predicted to exist, but have not been observed. We have determined
their energies to be $35,648.5(1)\ $cm$^{-1}$ and $35,757(1)\
$cm$^{-1}$ respectively. The existence of these states at these
energies gives a plausible explanation for the unusually large
polarizabilities found in the first part of the experiment. The
polarizabilities of two odd-parity states have also been
determined and are reported here.

\section{Experimental method}
\label{Section:ExpMeth}

\begin{figure}
\includegraphics[width=3.25 in]{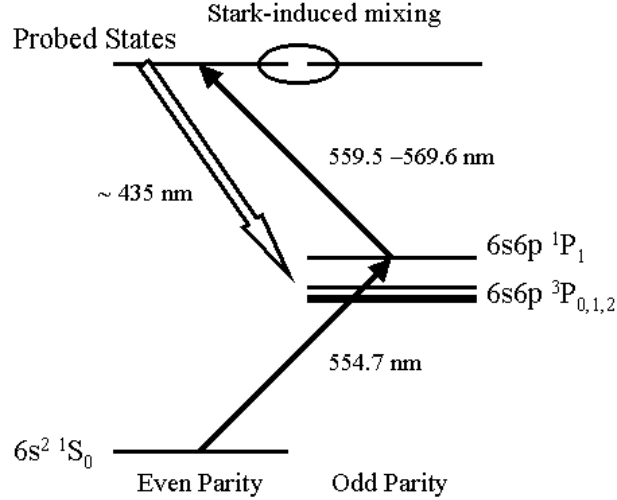}
\caption{The excitation-detection scheme. Solid arrows indicate
laser excitation; the hollow arrow indicates fluorescence. The
odd-parity states can be excited due to the Stark-induced mixing.}
\label{fig:excitation2}
\end{figure}

\begin{figure}
\includegraphics[width=3.25 in]{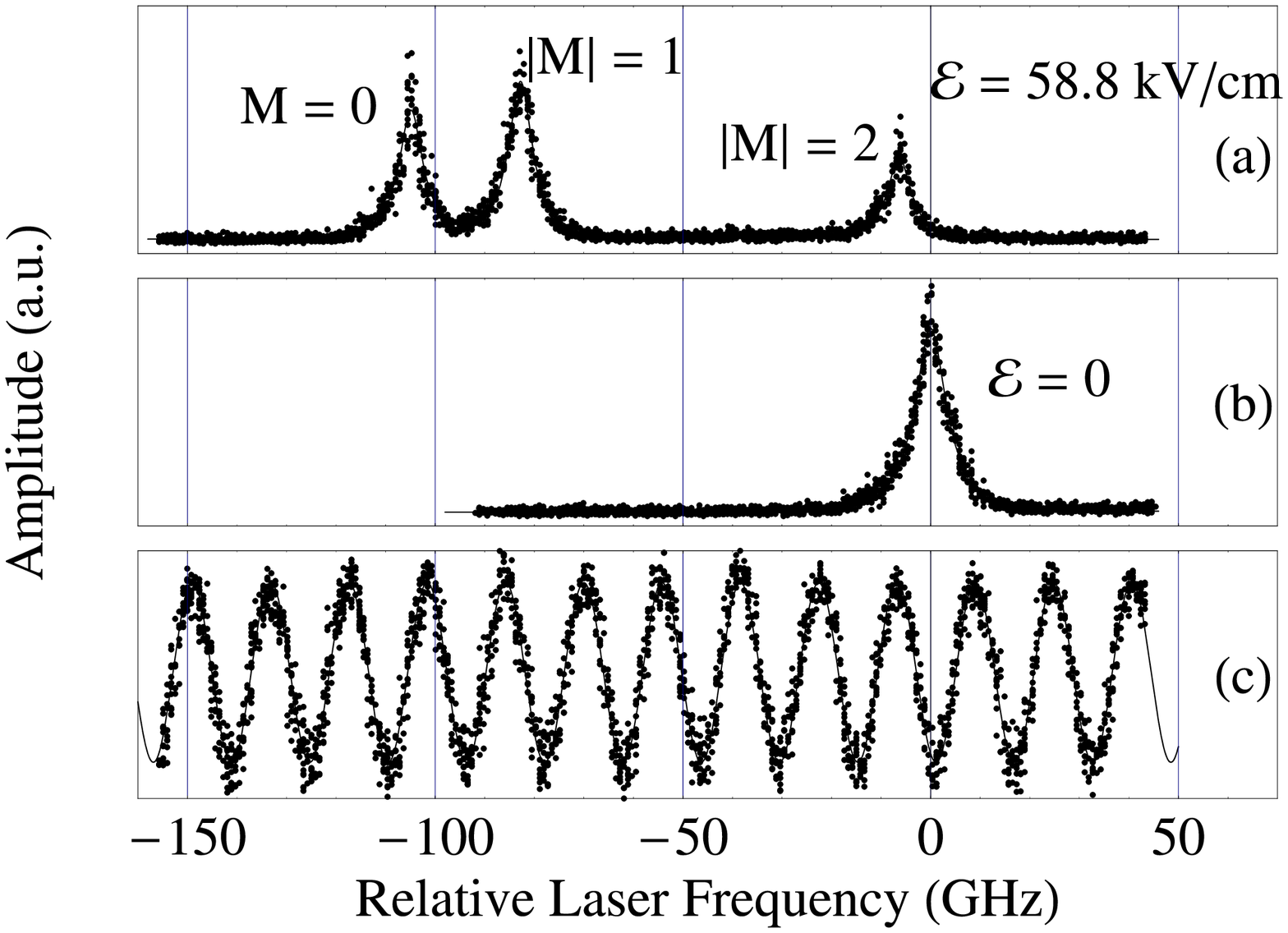}
\caption{The scan patterns for the 6p$^2\ ^3$P$_2$ state: (a)
amplitude of fluorescence with an electric field applied; (b)
amplitude of fluorescence with no electric field; (c) amplitude of
the reflection from the etalon used as a frequency marker. The
data points in (a) and (b) are fit by Lorentzian functions. The
data points in (c) are fit by a sinusoidal function.}
\label{fig:Ba6p6p3P2HV1}
\end{figure}

Pulsed lasers are used to excite barium atoms in an atomic beam to
the even-parity states of interest via two successive E1
transitions (Fig.~\ref{fig:excitation2}). Fluorescence resulting
from spontaneous decay to a lower-lying state is observed with a
photomultiplier tube (PMT). When an electric field is applied to
the interaction region, the resonance signal is Stark-shifted and
split (Fig.~\ref{fig:Ba6p6p3P2HV1}). In addition, Stark-induced
transitions to neighboring odd-parity states are seen; these are
also shifted and split by the electric field. The scalar and
tensor polarizabilities can be derived from appropriate fitting
functions relating the positions of the resolved resonance peaks
and the applied electric field.

\begin{figure}
\includegraphics[width=3.25 in]{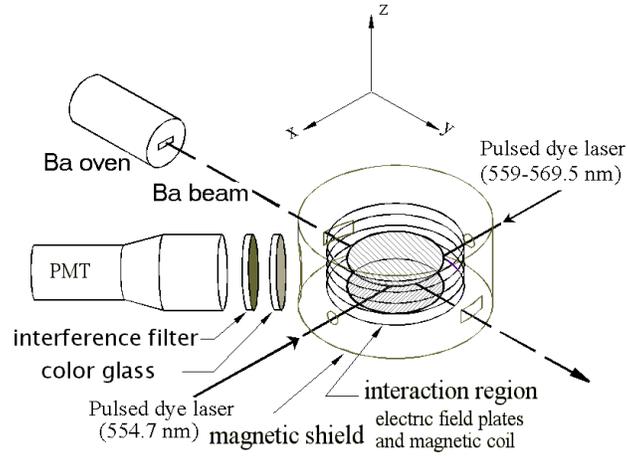}
\caption{Block diagram of the apparatus.} \label{fig:Chamber}
\end{figure}

The apparatus used (Fig.~\ref{fig:Chamber}) is largely the same as
in a previous experiment~\cite{Roc99}. The barium beam is produced
with an effusion source with a multi-slit nozzle that collimates
the angular spread of the beam to $\approx 0.1\ $rad in both the
horizontal and vertical directions. The oven, heat-shielded with
tantalum foil, is resistively heated to $\approx 700^{\circ}$C,
corresponding to saturated barium pressure in the oven of $\approx
0.1$~Torr and estimated atomic beam density in the interaction
region, $\approx 20$~cm away from the nozzle, of $\approx
10^{11}$~atoms/cm$^3$. Residual gas pressure in the vacuum chamber
is $\le 10^{-5}$~Torr.

The electric field is supplied by two plane-parallel electrodes
(angle $\le 10^{-3}$~rad) with diameter $6.4$~cm, spaced
$0.9153(3)$~cm apart. Voltage of up to $80$~kV is applied to the
top electrode using a high-voltage feed-through whose design is
described in detail in Ref.~\cite{Roc99}.

The laser systems in the experiment are two tunable dye lasers
(Quanta Ray PDL-$2$) pumped by pulsed frequency-doubled Nd-YAG
lasers (Quanta Ray DCR-$11$ and Quantel YAG$580$). The Quantel
laser operates at a repetition rate of $\approx 10$~Hz, and slaves
the Quanta Ray YAG laser, so the relative timing of the two
$\approx 10$-ns laser output pulses can be controlled to within
$\approx 1$~ns. In order to avoid shifts of the second-stage
transition spectra due to the dynamic Stark effect, the relative
timing of the two laser pulses is offset by $\approx 25$~ns. The
dyes used are Fluorescein~548 and Rhodamine~6G. For the dye laser
with Fluorescein~548, the wavelength is set at 554.7~nm (on
resonance with the 6s$^2\ ^1$S$_0$ $\rightarrow$ 6s6p~$^1$P$_1$
transition). The linewidth is $\approx 20$~GHz and the output is
linearly polarized parallel to the static electric field. The
typical pulse energy is $\approx 1.5$~mJ. For the dye laser with
Rhodamine~$6$G, the wavelength is coarsely tuned from 556~nm to
570~nm by tilting the diffraction grating in the laser cavity, and
finely tuned in a range of $\approx 0.2$~nm near a given resonance
by varying the gas (N$_2$) pressure in the laser-oscillator-cavity
chamber. The laser line width is $\approx 20$~GHz, or $\approx
6$~GHz when a narrowing etalon is employed. Typical pulse energy
is $3$~mJ. Color glass filters are used to reduce the pulse energy
to $\approx 0.4$~mJ to avoid power broadening effects. Both laser
beams are $\approx 3$~mm in diameter.

The two laser beams are sent into the chamber in anti-parallel
directions. To block scattered light from entering the chamber,
two $38$-cm-long collimating arms with several knife-edge
diaphragms are used for the entrance and exit of the laser beams.

Fluorescence is detected at $45^{\circ}$ to both the atomic and
laser beams with a $2$-in.-diameter PMT (type EMI 9750B). The gain
of the PMT is $\approx 7 \times 10^5$ and the quantum efficiency
at the wavelengths used is $\approx 25\%$. Interference filters
with $10$-nm bandwidth are used to select the decay channel of
interest, and a color glass filter is used to further reduce the
scattered light from the lasers.

In the first part of this experiment, we used CAMAC modules
connected through a general purpose interface bus (GPIB) to a
personal computer running LABVIEW software for data acquisition.
The fluorescence signal, as well as laser diagnostics, including
frequency markers, are recorded. In the second part, we observed
the time-dependent fluorescence signals with a digital
oscilloscope and determined the resonance wavelengths of the
second-stage laser from the reading of the grating position.

The frequency of the first-stage laser is set on resonance with
the 6s$^2\ ^1$S$_0$ to 6s6p~$^1$P$_1$ transition. We scan the
frequency of the second-stage laser across the resonance of a
certain probed state with total angular momentum $J$. When no
electric field is applied, we expect to see one resonance peak (or
nothing in the case of Stark-induced transitions). The shape of
the peak is mostly determined by the laser spectrum averaged over
many pulses and can be approximated with a Lorentzian. When the
electric field is applied along the z-axis, we expect the
resonance to shift and split into $J+1$ peaks (for the present
case of integer $J$), corresponding to sublevels with different
$|M|$, which we determined by varying the polarizations of both
laser beams, observing the variations of the resonant signal
amplitudes and comparing them to our expectations as following.

If the first-stage laser beam is linearly polarized along the
z-axis, the atoms can be excited to the $M=0$ sublevel of the
$^1$P$_1$ intermediate state. If the second-stage laser beam is
polarized along the z-axis, it can further excite the atoms to the
$M=0$ sublevel of a $J=0$ or $J=2$ probed state. Notice that the
transition to a $J=1\ M=0$ sublevel is forbidden because the
Clebsch-Gordan coefficient is zero. If the polarization of the
second-stage laser beam is along the y-axis, the atoms can be
further excited to the $|M|=1$ sublevels of a $J=1$ or $J=2$
probed state.

If the first-stage laser beam is linearly polarized along the
y-axis, the atoms can be excited to a coherent superposition of
$M=1$ and $M=-1$ sublevels ($(|M=1\rangle-|M=-1\rangle)/\sqrt{2}$)
of the $^1$P$_1$ intermediate state. If the second-stage laser
beam is polarized along the z-axis, it can further excite the
atoms to the $|M|=1$ sublevels of a $J=1$ or $J=2$ probed state.
If it is polarized along the y-axis, it can further excite the
atoms to the $M=0$ sublevel of a $J=0$ probed state or the $M=0$
and $|M|=2$ sublevels of a $J=2$ probed state. Notice that the
transition to a $J=1\ M=0$ sublevel is forbidden because the
transition amplitudes from the $M=1$ and $M=-1$ sublevels of the
$^1$P$_1$ state have the same magnitude but opposite signs so that
the sum of the amplitudes is zero. The relation between the
polarizations of the laser beams and the excited Zeeman sublevels
of a state with total angular momentum $J$ is summarized in
Table~\ref{table:laserpol}.

\section{Theory of polarizability}
\label{Section:theory}

In this section, we briefly introduce some definitions and
formulae used in this work. For a detailed discussion, see Refs.
\cite{Kul84,Sobelman}.

The Hamiltonian for an atom in a DC electric field
$\vec{E}=\mathcal{E}\hat{z}$ can be written as:
\begin{equation}
H=H_0-\vec{E}\cdot\vec{D}=H_0-\mathcal{E D}_z,
\label{Eqn:Hamiltonian}
\end{equation}
where $H_0$ is the Hamiltonian in the absence of electric field,
with eigenenergies and eigenstates $E_{a_0 J_0 M}$ and $|a_0 J_0
M\rangle$, where $a_0$ is electronic configuration, $J_0$ is total
electronic angular momentum, $M$ is its z-component, $\vec{D}=-e
\vec{r}$ is the electric dipole operator, and $e$ is the magnitude
of the electron charge. If there are only two states, exact
eigenenergies can be obtained from the secular equation:
\begin{equation}
\left| \begin{array}{cc}
-\delta_S E_{a_0 J_0 M} & -\mathcal{E D}_z\\
-\mathcal{E D}_z & -\Delta E-\delta_S E_{a_0 J_0 M}
\end{array} \right|=0,
\label{Eqn:secular}
\end{equation}
where $\Delta E= E_{a_0 J_0 M}-E_{a_1 J_1 M}$ and
$\mathcal{D}_z=\langle a_1 J_1 M|D_z|a_0 J_0 M\rangle$
($M_1=M_0=M$ for z-directed electric field). From
Eq.~(\ref{Eqn:secular}), the Stark energy shift of state $|a_0 J_0
M\rangle$ is,
\begin{equation}
\delta_S E_{a_0 J_0 M}=-\frac{\Delta E}{2} \pm
\sqrt{\prn{\frac{\Delta E}{2}}^2 + \mathcal{D}_z^2 \mathcal{E}^2},
\label{Eqn:2leveldE}
\end{equation}
where the sign on the right-hand side of the equation should be
chosen the same as the sign of $\Delta E$. If the electric field
is weak, so that $\mathcal{D}_z^2 \mathcal{E}^2 \ll \Delta E^2/4$,
Eq.~(\ref{Eqn:2leveldE}) can be expanded in series. To the lowest
order,
\begin{equation}
\delta_S E_{a_0 J_0 M}^{(2)}=\frac{\mathcal{D}_z^2
\mathcal{E}^2}{\Delta E}=-\frac{\alpha_{a_0 J_0
M}}{2}\mathcal{E}^2, \label{Eqn:2levelsmallEdE}
\end{equation}
where $\alpha$ is the static polarizability. Thus in a two-level
system, one can express the polarizability as:
\begin{equation}
\alpha_{a_0 J_0 M}=-2 \frac{\mathcal{D}_z^2}{\Delta E}.
\label{Eqn:2-level alpha}
\end{equation}
If there is more than one state coupled to a state $|a_0 J_0 M
\rangle$, to the lowest order, the energy shift is just a sum over
all contributions:
\begin{equation}
\delta_S E_{a_0 J_0 M}^{(2)}=\sum_{a_1 J_1}\frac{\mathcal{D}_z^2
\mathcal{E}^2}{E_{a_0 J_0 M}-E_{a_1 J_1 M}}, \label{Eqn:2order
purt}
\end{equation}
so that
\begin{equation}
\alpha_{a_0 J_0 M}=-2\sum_{a_1 J_1}\frac{\mathcal{D}_z^2}{E_{a_0
J_0 M}-E_{a_1 J_1 M}}. \label{Eqn:alpha of multilevel}
\end{equation}
The polarizability is usually expressed in terms of its scalar and
tensor parts as:
\begin{equation}
\alpha_{a_0 J_0 M}=\alpha_{0,a_0 J_0}+\alpha_{2,a_0
J_0}\frac{3M^2-J_0(J_0+1)}{J_0(2J_0-1)}. \label{Eqn:alpha0alpha2}
\end{equation}
The scalar polarizability, $\alpha_0$, represents the average
shift of different $M$ sublevels; the tensor polarizability,
$\alpha_2$, represents the differential shift. For states with $J
\le 1/2$, there is no Stark-induced splitting and, therefore,
$\alpha_2$ is zero.

Using the formalism of the reduced matrix elements,
\begin{eqnarray}
        \langle a_1 J_1 M_1|D_q|a_0 J_0 M_0\rangle &=& (-1)^{J_1-M_1}
        \left( \begin{array}{ccc} J_1 & 1 & J_0 \\ -M_1 & q & M_0
        \end{array} \right)
        \langle a_1 J_1||D||a_0 J_0\rangle,
\label{a4a}
\end{eqnarray}
the scalar and tensor polarizabilities can be expressed as:
\begin{eqnarray}
        \alpha_{0,a_0 J_0} &=& \frac{-2}{3(2J_0+1)}
        \sum_{a_1 J_1} \frac{|\langle a_0 J_0||D||a_1 J_1\rangle|^2} {E_{a_0 J_0 M_0}-E_{a_1 J_1 M_1}},
\label{a3}
\\
        \alpha_{2,a_0 J_0} &=&
        4 \left(\frac{5J_0(2J_0-1)}{6(2J_0+3)(2J_0+1)(J_0+1)}\right)^{1/2}\cdot
        \\
 &&       \sum_{a_1 J_1} (-1)^{J_0+J_1+1}
        \left\{ \begin{array}{ccc} J_0 & 1 & J_1 \\ 1 & J_0 & 2
        \end{array} \right\}
        \frac{|\langle a_0 J_0||D||a_1 J_1\rangle|^2} {E_{a_0 J_0 M_0}-E_{a_1 J_1
        M_1}}.
\label{a4}
\end{eqnarray}

Using Eq.(\ref{Eqn:alpha0alpha2}), $\alpha_{a_0 J_0 M}$ can also
be expressed as:
\begin{equation}
\alpha_{a_0 J_0 M}=A_{a_0 J_0}+B_{a_0 J_0}M^2, \label{Eqn:alphaAB}
\end{equation}
where $A_{a_0 J_0}=\alpha_{0,a_0 J_0}-\alpha_2(J_0+1)/(2J_0-1)$
and $B_{a_0 J_0}=3\alpha_2/[J_0(2J_0-1)]$. In this expression, it
is clearly seen that the difference between the Stark-induced
energy shifts of $|M|$ sublevels and that of $M=0$ sublevel is
proportional to $M^2$ when the energy shifts are quadratically
dependent on the electric field. In the data analysis, this
relation provides us with a way to determine $|M|$ for each
Stark-split resonance peak for states with $J\ge2$.

When the applied electric field is higher, but still low enough so
that the perturbation theory is valid, one needs to keep the next
(fourth) order terms in the expansion:
\begin{equation}
\delta_S E_{a_0 J_0 M}=\delta_S E_{a_0 J_0 M}^{(2)}+\delta_S
E_{a_0 J_0 M}^{(4)}+..., \label{Eqn:4expansion}
\end{equation}
where $\delta_S E_{a_0 J_0 M}^{(2)}$ is given by
Eq.(\ref{Eqn:2order purt}), and
\begin{equation}
\begin{split}
\delta_S E_{a_0 J_0 M}^{(4)}=&\sum_{a_1 J_1 a_2 J_2 a_3 J_3
}\frac{\mathcal{D}_z^{01}\mathcal{D}_z^{12}\mathcal{D}_z^{23}\mathcal{D}_z^{30}
\mathcal{E}^4}{(E_{a_0 J_0 M}-E_{a_1 J_1 M})(E_{a_0 J_0 M}-E_{a_2
J_2 M})(E_{a_0 J_0 M}-E_{a_3 J_3 M})}\\
&-\sum_{a_1 J_1 a_2 J_2}\frac{|\mathcal{D}_z^{01}|^2
|\mathcal{D}_z^{02}|^2 \mathcal{E}^4}{(E_{a_0 J_0 M}-E_{a_1 J_1
M})^2 (E_{a_0 J_0 M}-E_{a_2 J_2 M})}, \label{Eqn:4perturbation}
\end{split}
\end{equation}
where $\mathcal{D}_z^{ij}=\langle a_i J_i M|D_z|a_j J_j M\rangle$.
The hyperpolarizability is defined by:
\begin{equation}
\delta_S E_{a_0 J_0 M}^{(4)}=-\frac{\gamma_{a_0 J_0
M}}{4!}\mathcal{E}^4. \label{Eqn:4order}
\end{equation}
The hyperpolarizability is usually expressed in terms of the
scalar and tensor hyperpolarizabilities, $\gamma_0$ and
$\gamma_2$, $\gamma_4$ as:
\begin{equation}
\begin{split}
\gamma_{a_0 J_0 M}=&\gamma_{0,a_0 J_0}+\gamma_{2,a_0
J_0}\frac{3M^2-J_0(J_0+1)}{J_0(2J_0-1)}\\
&+\gamma_{4,a_0
J_0}\frac{35M^4+[25-30J_0(J_0+1)]M^2+(J_0-1)J_0(J_0+1)(J_0+2)}{J_0(2J_0-1)(2J_0-2)(2J_0-3)}.
\label{Eqn:gamma}
\end{split}
\end{equation}
For states with $J\le 1/2$, $\gamma_2$ and $\gamma_4$ are zero.
For states with $J\le 3/2$, $\gamma_4$ is zero. The derivation of
Eq.~(\ref{Eqn:4perturbation}) and Eq.~(\ref{Eqn:gamma}) can be
found in Ref. \cite{Kul84}.

When a weak external z-directed electric field is applied, to the
lowest order, the perturbed state is mixed with opposite-parity
states according to:
\begin{equation}
|``a_0 J_0 M"\rangle=|a_0 J_0 M\rangle+\sum_{a_1
J_1}\frac{\mathcal{D}_z \mathcal{E}}{E_{a_0 J_0 M}-E_{a_1 J_1 M}}\
|a_1 J_1 M\rangle. \label{Eqn:statemixing}
\end{equation}
For example, a small portion of a nominally odd-parity state has
even parity. The mixing amplitude is proportional to the applied
electric field. Therefore, a nominally odd-parity state can be
excited from the ground state via two E$1$ transitions when the
electric field is applied. The transition probability to this
state, which is proportional to the square of the mixing
amplitude, is quadratically dependent on the applied electric
field. In the experiment, we detect the Stark-induced signal from
the decay of the even-parity component of the nominally odd-parity
state to 6s8p~$^3$P$_1$, which makes the emission probability also
quadratically dependent on the electric field. Therefore, for weak
electric fields, the amplitude of the signal scales as the fourth
power of the field. On the other hand, if the electric field is so
strong that the states of opposite parity are fully mixed, the
signal amplitude is almost independent of the applied electric
field. Thus, the dependence of the signal amplitude on the applied
electric field can be used to check the validity of the
perturbation approximation.

\section{Results and discussion}
\label{Section:ResultsDisc}

This section is divided into two subsections. In the first part,
we discuss the Stark effect in four even-parity states ranging
from $35,600$ to $36,000$~cm$^{-1}$. The polarizabilities of three
of these states are too large to be explained by the known energy
spectrum of barium and imply the existence of ``new" states. As
shown in Fig.~\ref{fig:SmallRangeEd}, within the energy range
probed in this experiment, one may expect the $J=0$ and $J=2$ of
6s8p~$^3$P$_J$ states that have not been identified before are
candidates for these new states. In the second part, we discuss
the observation of the Stark-induced transitions to the odd-parity
states. Two of them cannot be identified with the known energy
levels of barium. They are possibly the 6s8p~$^3$P$_{0,2}$ states
and their energies are determined in this experiment. With the
existence of these two ``new" states, the unusually large
polarizabilities found in the first subsection can be plausibly
explained.

The measured relative energy shifts include the Stark-induced
shift of the intermediate state, 6s6p~$^1$P$_1$. However, the
scalar and tensor polarizabilities of the 6s6p~$^1$P$_1$ state are
$51(3)$~kHz/(kV/cm)$^2$ and
$-5.34(5)$~kHz/(kV/cm)$^2$~\cite{Koz99}, respectively, and
contribute to the shift at a level $\le 200$~MHz, negligible
compared to the observed shifts.

\subsection{Stark effect of even-parity states}

The derived scalar and tensor polarizabilities of 6p$^2\ ^3$P$_2$,
6s7d~$^3$D$_{1,2}$ and 5d6d~$^3$D$_1$ states are listed in
Table~\ref{table:polarizability}. Multiple resonance peaks are
observed for each of these four states corresponding to different
values of $|M|$. Moreover, the Stark shifts of two states show
significant deviation from quadratic behavior as a function of the
electric field, so that the scalar and tensor
hyperpolarizabilities can be measured
(Table~\ref{table:hyperpolarizability}). The detailed experimental
results for these four states are discussed in the following.

\paragraph{The {\rm 6p$^2\ ^3$P$_2$} state.}

\begin{figure}
\includegraphics[width=3.25 in]{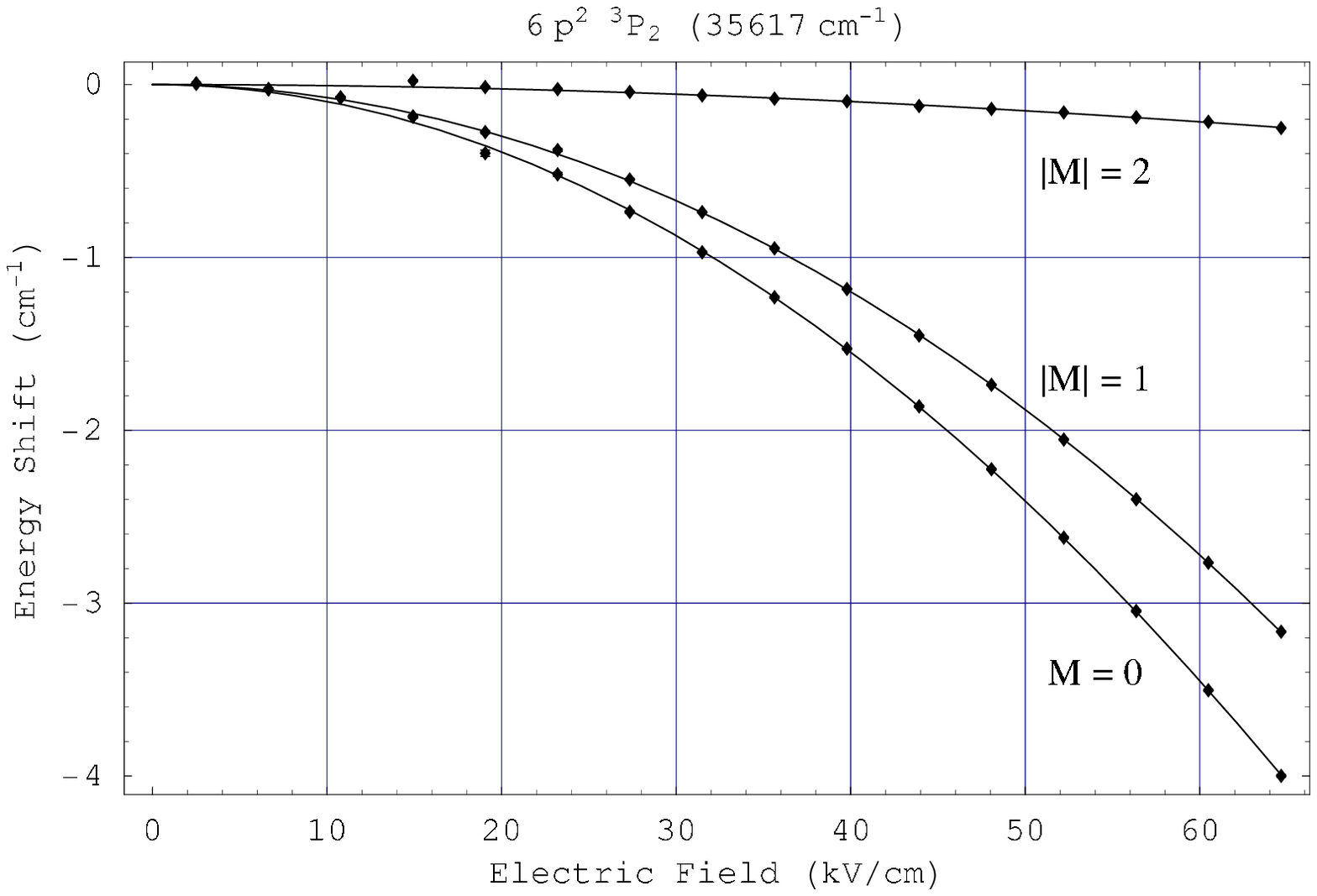}
\caption{The Stark splitting and shift for the resonance of the
6p$^2\ ^3$P$_2$ state. The data points are fit by
Eq.(\ref{Eqn:4expansion}).} \label{fig:Ba6p6p3P2}
\end{figure}

With an electric field applied, three resonance peaks are observed
for this state
(Figs.~\ref{fig:Ba6p6p3P2HV1},~\ref{fig:Ba6p6p3P2}). We locate the
centers of the peaks by fitting the data with Lorentzian
functions. The peaks are put into correspondence with Zeeman
sublevels using Eq.~(\ref{Eqn:alphaAB}) and their amplitude
dependence on laser polarization. Although the observed level
shifts are dominated by the polarizability, we are also able to
resolve the hyperpolarizability of each sublevel by fitting with
Eq.~(\ref{Eqn:4expansion}).

\begin{figure}
\includegraphics[width=3.25 in]{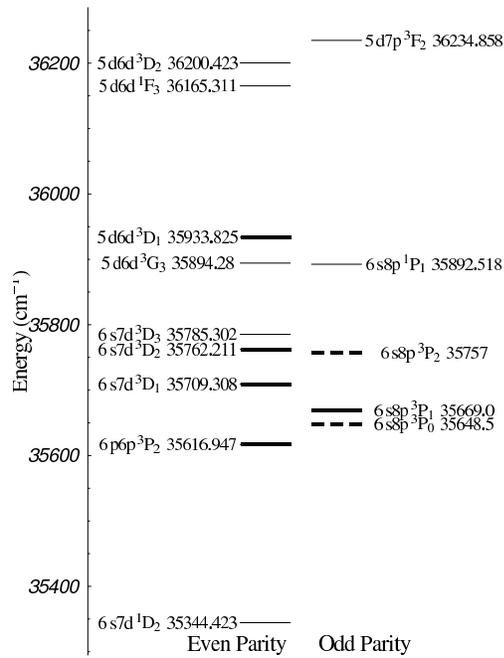}
\caption{The energy-level diagram of barium ranging from $35,300$
to $36,300$~cm$^{-1}$. The dashed lines are the suggested ``new"
energy levels. The seven bold lines are the states whose
polarizabilities are measured in this
work.}\label{fig:SmallRangeEd}
\end{figure}

The closest known state that can Stark-mix with the 6p$^2\
^3$P$_2\ M=0$ sublevel is 6s8p~$^3$P$_1$~\cite{Arm79}, with
$\Delta E=-52.0(2)\ $cm$^{-1}$ (Fig.~\ref{fig:SmallRangeEd}). If
we assume that this is the dominantly coupled state, from the
shift of the $M=0$ sublevel, we find for the reduced electric
dipole moment: $|(6$p$^2\ ^3$P$_2\|D\|6$s8p~$^3$P$_1)_{\rm
two-level}|=14.7(2)$~ea$_0$ (Table~\ref{table:dipole}). This is
far too large, considering that the electronic configurations of
these two states are different by two electrons. Thus we can
conclude that the coupling to other states and/or the effects of
configuration mixing are not negligible.

The closest known state that can couple to the 6p$^2\ ^3$P$_2$
$|M|=1$ sublevels is again $6$s$8$p~$^3$P$_1$. Using the two-level
approximation, we find that the reduced electric dipole moment is
$|(6$p$^2\ ^3$P$_2\|D\|6$s8p~$^3$P$_1)_{\rm two-level }|=
14.3(2)$~ea$_0$, which is again too large. However, the fact that
the obtained reduced electric dipole moments derived from $M=0$
and $|M|=1$ sub-levels are nearly the same may suggest that the
6p$^2\ ^3$P$_2$ state and the 6s8p~$^3$P$_1$ state indeed have
large electric dipole coupling perhaps due to admixtures of other
configurations.

There is no apparent dominant close energy state that can couple
to the $|M|=2$ sublevels. This explains the fact that the shift of
the $|M|=2$ sublevels is an order of magnitude smaller than that
of $|M|=1$ and $M=0$ sublevels.

\paragraph{The {\rm 6s7d~$^3$D$_1$} state.}

\begin{figure}
\includegraphics[width=3.25 in]{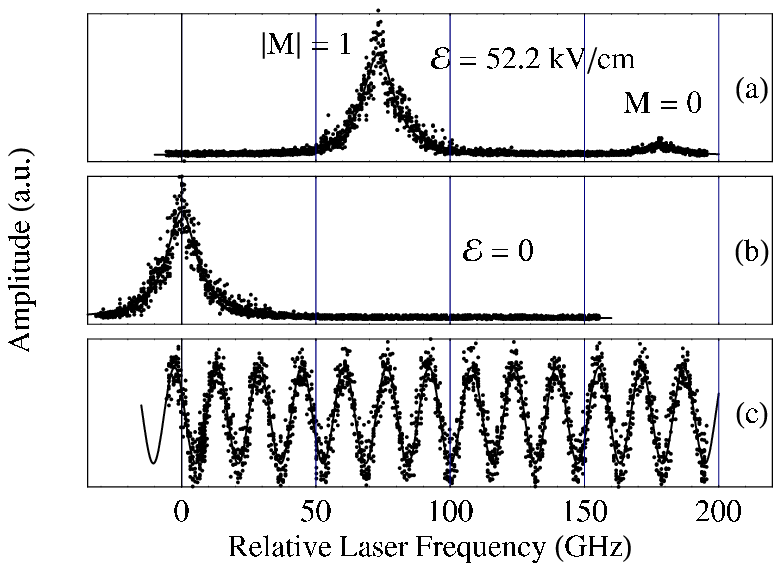}
\caption{The scan patterns for the 6s7d~$^3$D$_1$ state: (a)
amplitude of fluorescence with an electric field applied; (b)
amplitude of fluorescence with no electric field; (c) amplitude of
the reflection from the etalon used as a frequency marker. The
data points in (a) and (b) are fit by Lorentzian functions. The
data points in (c) are fit by a sinusoidal function.}
\label{fig:Ba6s7d3D1HV1}
\end{figure}

\begin{figure}
\includegraphics[width=3.25 in]{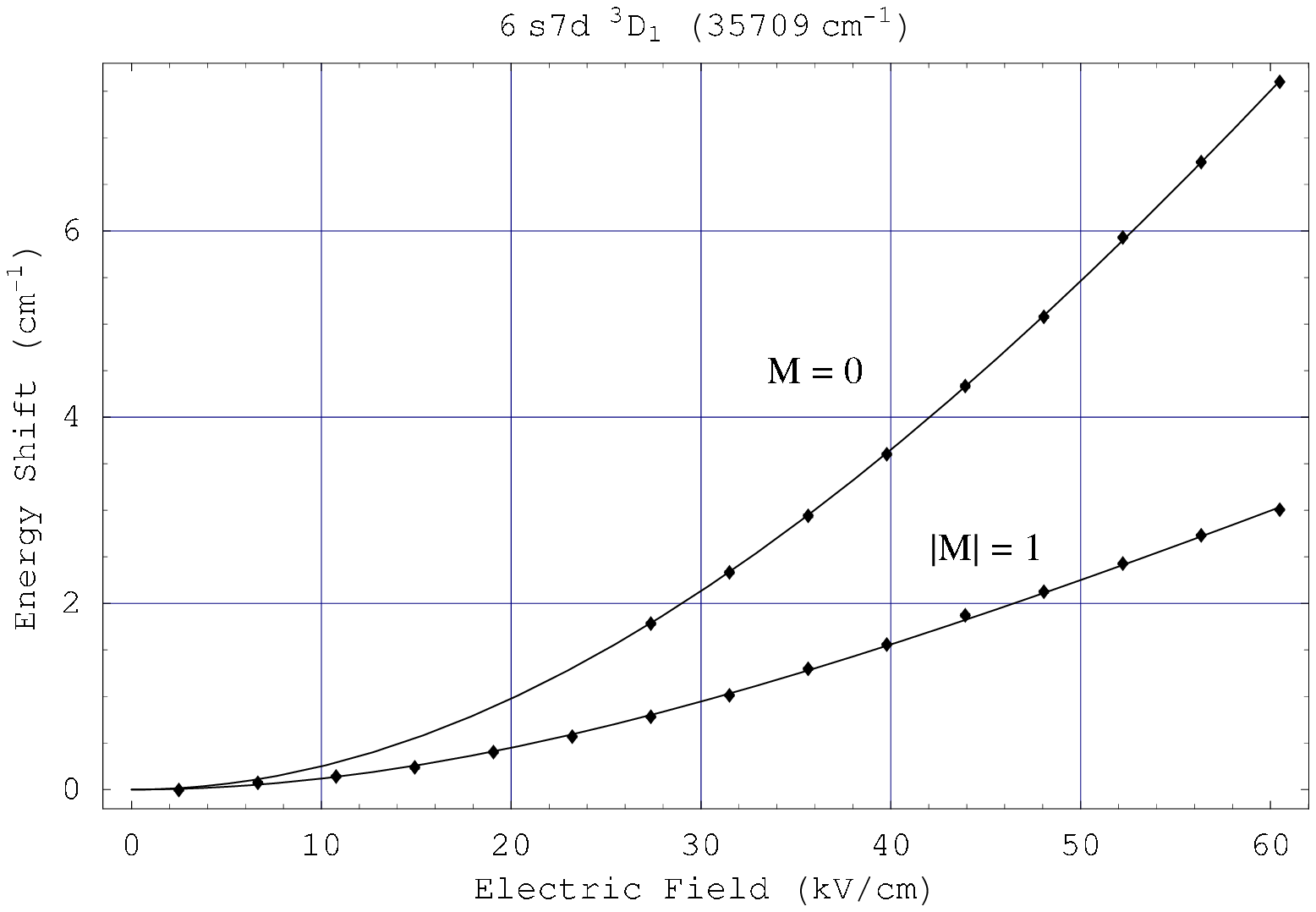}
\caption{The Stark splitting and shift for the resonance of the
6s7d~$^3$D$_1$ state. The data points are fit by
Eq.(\ref{Eqn:4expansion}).} \label{fig:Ba6s7d3D1}
\end{figure}

With electric field applied, the resonance peak is split into two
peaks, corresponding to the $M=0$ and $|M|=1$ sublevels,
respectively (Figs.~\ref{fig:Ba6s7d3D1HV1},~\ref{fig:Ba6s7d3D1}).
The size of the resonance peak corresponding to the
$|J=1~M=0\rangle$ sublevel is more than one order of magnitude
smaller because in our experimental setup, the transition to this
sublevel is nominally forbidden. It is still observed most likely
because of the imperfections in the direction and purity of the
laser polarizations.

Neither of the curves in Fig.~\ref{fig:Ba6s7d3D1} can be
adequately fit with quadratic functions at high electric fields.
Therefore, the next (fourth) order terms in the perturbation
expansion should be included in the fitting function and both
polarizabilities and hyperpolarizabilities can be determined.

If there is a dominantly coupled state for each of the $|M|$
sublevels, the Stark energy shifts can be described by the exact
solution of the two-level Hamiltonian, Eq.~(\ref{Eqn:2leveldE}).
Under this two-level approximation, the energy difference and the
electric dipole couplings between the $|M|$ sublevels and their
dominant coupling state can be determined separately, and the
polarizabilities can be obtained from Eq.~(\ref{Eqn:2-level
alpha}).

The large polarizability of the 6s7d~$^3$D$_1$ $|M|=1$ sub-levels
might have been plausibly explained by the coupling with the close
$6$s$8$p~$^3$P$_1$ state. However, under the two-level
approximation, we obtain $\Delta E|_{\rm two-level}\approx
7$~cm$^{-1}$, which is about five times smaller than the actual
energy difference of 6s7d~$^3$D$_1$ and 6s8p~$^3$P$_1$ states
\cite{Arm79}. This shows that the two-level approximation breaks
down and the mixing with other states is not negligible.

The large polarizability of 6s7d~$^3$D$_1$~$M=0$ sublevel cannot
be explained by the known odd-parity energy levels. Using the
two-level approximation, one odd-parity state with $J=0$ or $J=2$
is predicted to be at $\approx 35,673$~cm$^{-1}$. However, it is
shown in Fig.~\ref{fig:SmallRangeEd} that 6s8p~$^3$P$_1$ is the
only odd-parity state known with that energy. As will be discussed
in the next subsection, no unidentified ``new" state is expected
there either. This shows that two-level approximation is not
applicable. Nevertheless, there should be at least one $J=0$ or
$J=2$ odd-parity state lying somewhat below $35,709$~cm$^{-1}$,
the energy of 6s7d~$^3$D$_1$ state, in order to explain the large
negative polarizability of this sublevel. This prediction is
confirmed in the next subsection.

\paragraph{The {\rm 5d6d~$^3$D$_1$} state.}

\begin{figure}
\includegraphics[width=3.25 in]{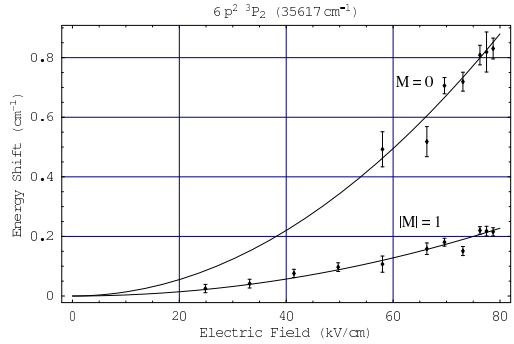}
\caption{The Stark splitting and shift for the resonance of the
5d6d~$^3$D$_1$ state. The data points are fit by quadratic
functions.} \label{fig:Ba5d6d3D1}
\end{figure}

With the electric field applied, the resonance peak of the
5d6d~$^3$D$_1$ state is split into two peaks, corresponding to
$|M|=1$ and $M=0$ sublevels (Fig.~\ref{fig:Ba5d6d3D1}). The
amplitude of the resonance peak of $M=0$ sublevel is, as expected,
much smaller than that of the $|M|=1$ sublevels. For both curves
in Fig.~\ref{fig:Ba5d6d3D1}, quadratic functions are used in the
fitting to determine polarizabilities. The polarizabilities of
this state are at least one order of magnitude smaller than others
and agree with our estimate based on the known energy levels of
barium.

\paragraph{The {\rm 6s7d~$^3$D$_2$} state.}

\begin{figure}
\includegraphics[width=3.25 in]{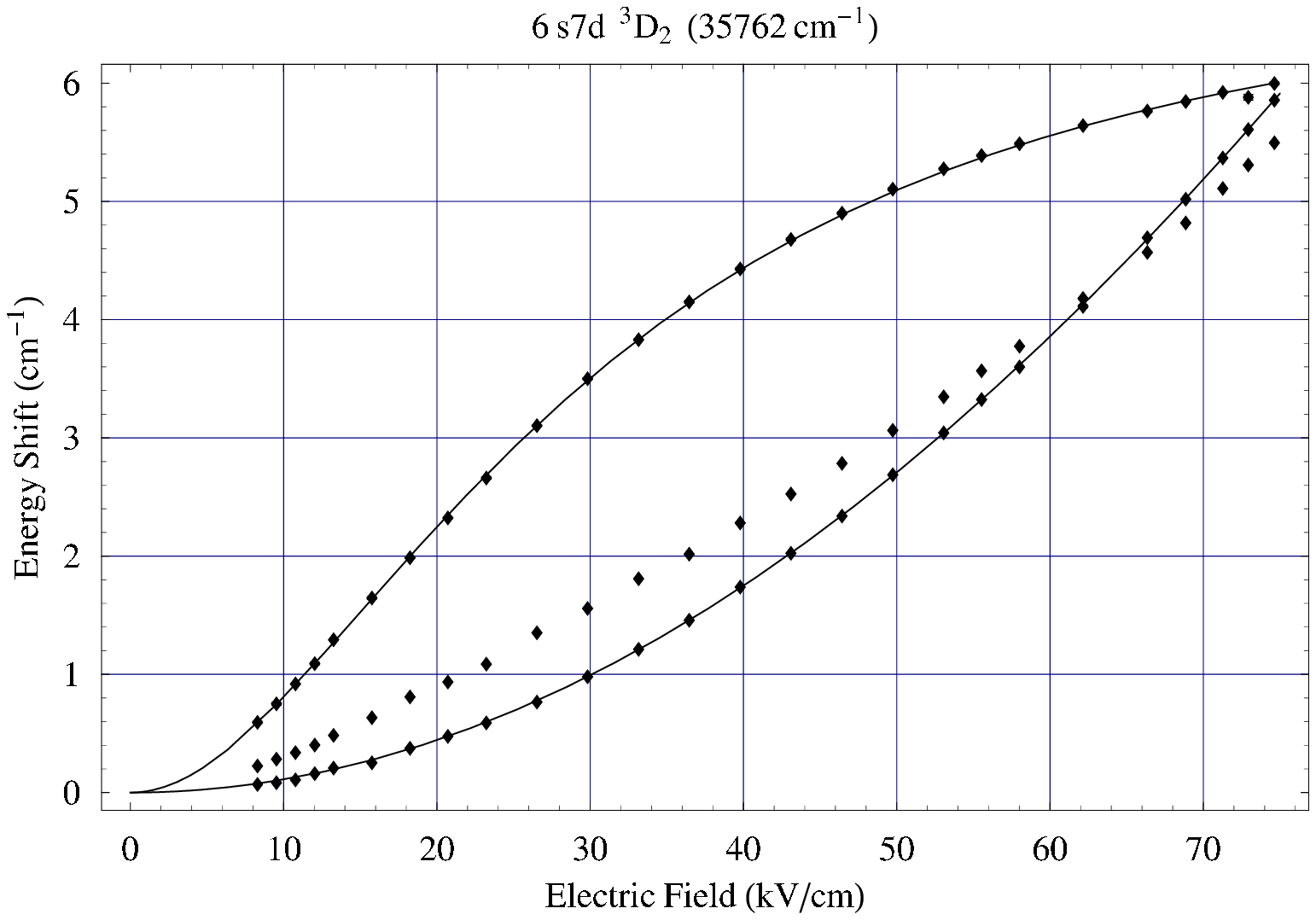}
\caption{The Stark splitting and shift for the resonance of the
6s7d~$^3$D$_2$ state. The data points corresponding to $M=0$
sublevel are fit by the exact solutions of the two-level
Hamiltonian. The data points corresponding to $|M|=2$ sub-levels
are fit by the exact solutions of the three-level Hamiltonian. The
data points corresponding to $|M|=1$ sub-levels cannot be fit by
quadratic functions or modelled by a two- or three-level system.}
\label{fig:Ba6s7d3D2}
\end{figure}

The plot of the Stark-induced energy shift of the 6s7d~$^3$D$_2$
state as a function of the applied electric field is shown in
Fig.~\ref{fig:Ba6s7d3D2}. For the $M=0$ sublevel, a quadratic
function is not adequate to fit the data points and the fourth
order term is included in the fitting function. With
Eq.~(\ref{Eqn:4perturbation}), we determined the polarizability
$\alpha(M=0)=-66.6(5)$~MHz/(kV/cm)$^2$ and hyperpolarizability
$\gamma(M=0)=8(2)$~kHz/(kV/cm)$^4$.

\setcounter{footnote}{0}

The 6s7d~$^3$D$_2$~$M=0$ sublevel can be expected to dominantly
couple to the 6s8p~$^3$P$_1$~$M=0$. The coupling to 6s8p~$^1$P$_1$
is suppressed because of the difference in total
spins\footnote{The spin-orbit interaction can mix the
6s8p~$^3$P$_1$ and 6s8p~$^1$P$_1$ states so that the electric
dipole coupling between the 6s7d~$^3$D$_2$ and 6s8p~$^1$P$_1$
states is not zero. This mixing, which is estimated in section V,
is negligible in present crude estimation.}. Under the two-level
approximation, the derived $\Delta E|_{\rm two-level
}=103(20)$~cm$^{-1}$ is consistent with the energy difference
between those two states, $93.2(2)$~cm$^{-1}$. The derived reduced
electric dipole moment $|(6$s7d~$^3$D$_2\|D\|6$s8p~$^3$P$_1)_{\rm
two-level }|= 22(2)$~ea$_0$ is not too different from
$34.7$~ea$_0$, that we estimate using the Bates-Damgaard
approximation using expressions given in~\cite{Sobelman}.

For the $|M|=1$ and $|M|=2$ sublevels (as most clearly seen in
Fig.~\ref{fig:Ba6s7d3D2}, for $|M|=2$) the Stark-induced energy
shifts do not obey the electric-field dependence of
Eq.~(\ref{Eqn:4expansion}) and the perturbation theory is not
applicable. The enormous shifts of all sublevels require an
odd-parity state with $J=2$ or $3$, with energy slightly lower
than $35,762.211$~cm$^{-1}$, the energy of 6s7d~$^3$D$_2$ state.
However, it seems that the $M=0$ sublevel is not coupled to this
new state; otherwise the two-level approximation would not apply
for $M=0$ sublevel. Therefore, we expect this new state with
$J=2$. Another constraint on the term of this new state is that
the electric dipole moment between this $J=2$ state and the
6s7d~$^3$D$_1$ state should be small compared with
(6s7d~$^3$D$_1\|D\|6$s8p~$^3$P$_{0,1})$; otherwise it would
strongly cancel the contribution of the polarizabilities of the
6s7d~$^3$D$_1$ state from the 6s8p~$^3$P$_{0,1}$ states. It is
known that the electric dipole moment between two states with
$\Delta J=-\Delta L$, where $L$ is the orbital angular momentum,
is suppressed ~\cite{Sobelman} (See also Table V). Therefore, on
the basis of all these arguments, this new state is predicted to
be a $^3$P$_2$ state. This prediction is confirmed by the
observation of one of the Stark-induced transitions discussed in
the next subsection. The missing state is most likely
6s8p~$^3$P$_2$.

The Stark effect of the $|M|=2$ sublevels could be modelled with a
three-level system, consisting of 6s7d~$^3$D$_3$, 6s7d~$^3$D$_2$,
and 6s8p~$^3$P$_2$. Notice that although $|6$s7d~$^3$D$_2\
|M|=2\rangle$ sublevels only dominantly couple to
$|6$s8p~$^3$P$_2\ |M|=2\rangle$, the two-level approximation is
not adequate in this case, because $|6$s8p~$^3$P$_2\ |M|=2\rangle$
is also strongly coupled to $|6$s7d~$^3$D$_3\ |M|=2\rangle$. The
couplings to the close-lying states other than these three are
suppressed because of the difference in total spins or the
different configuration of both electrons. With the three-level
approximation, we have determined the energy of 6s8p~$^3$P$_2$
state as $35,756(1)$~cm$^{-1}$ and obtained two reduced electric
dipole moments, $|(6$s7d~$^3$D$_2\|D\|6$s8p~$^3$P$_2)_{\rm
three-level }|= 16.2(5)$~ea$_0$ and
$|(6$s7d~$^3$D$_3\|D\|6$s8p~$^3$P$_2)_{\rm three-level}|=
40(2)$~ea$_0$. The polarizability of the $|M|=2$ sublevels is
found to be $\alpha(|M|=2)=-881(13)$~MHz/(kV/cm)$^2$. Knowing
$\alpha(|M|=2)$ and $\alpha(M=0)$, in the three-level
approximation, we obtain $\alpha_0=-335(13)$~MHz/(kV/cm)$^2$ and
$\alpha_2=-268(13)$~MHz/(kV/cm)$^2$.

\subsection{Observation of the Stark-induced transitions}

In this subsection, we discuss the observation of the
Stark-induced transitions to the odd-parity states. Within the
energy range studied, five Stark-induced transitions are detected
and classified into three sets. Each set of resonances is
discussed in detail in the following.

\begin{figure}
\includegraphics[width=3.25 in]{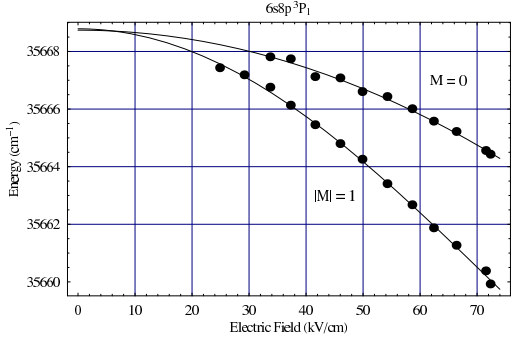}
\caption{The Stark splitting and shift for the resonance of the
6s8p~$^3$P$_1$ state. The data points corresponding to $|M|=1$
sub-levels are fit by a quadratic function. The data points
corresponding to $M=0$ sub-level are fit by
Eq.(\ref{Eqn:4expansion}).} \label{fig:Ba6s8p3P1}
\end{figure}

Two of the Stark-induced resonances are classified as the first
set. According to the electric-field dependence of the amplitudes
of the signals, the perturbation theory applies in both cases. The
resonance position as a function of the applied electric field is
shown in Fig.~\ref{fig:Ba6s8p3P1}. The fitting function for
$|M|=1$ sublevels is a quadratic function plus a constant term,
which corresponds to the energy of the zero-electric-field
resonance, which cannot be measured directly in our experiment.
For the $M=0$ sublevel, the fourth order term is necessary and
Eq.~(\ref{Eqn:4expansion}) plus a constant is employed as the
fitting function. The derived energies of the zero-electric-field
resonance from each curve are $35,668.7(2)$ and
$35,668.8(2)$~cm$^{-1}$, respectively. The coincidence of the two
values shows that they are the split Zeeman sublevels of the same
state. The fact that this derived energy coincides with the energy
of the 6s8p~$^3$P$_1$ state, $35,669.0(2)$~cm$^{-1}$ \cite{Arm79},
indicates that these two resonance peaks are the Stark-induced
transitions to the $|M|=1$ and $M=0$ sublevels of this state. The
$|M|$ of each curve is determined from the amplitude dependence on
laser polarization. From the electric-field dependence of the
resonance position, we have determined the polarizabilities of the
$6$s$8$p~$^3$P$_1$ state (Table~\ref{table:polarizability}) and
the hyperpolarizability $\gamma(M=0)=-47(11)$~kHz/(kV/cm)$^4$.

\begin{figure}
\includegraphics[width=3.25 in]{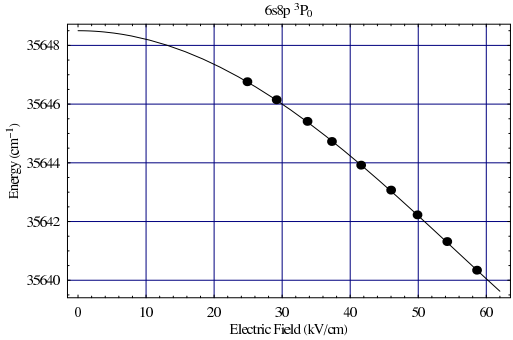}
\caption{The Stark shift for the resonance of the 6s8p~$^3$P$_0$
state. The data points are fit by Eq.(\ref{Eqn:4expansion}).}
\label{fig:Ba6s8p3P0}
\end{figure}

One of the Stark-induced resonances is classified as belonging to
a set by itself. According to the electric-field dependence of the
amplitude of the signal, the perturbation theory applies in this
case. The resonance position as a function of the applied electric
field is shown in Fig.~\ref{fig:Ba6s8p3P0}. The data points are
fit with Eq.~(\ref{Eqn:4expansion}) plus a constant term. The
polarizability can be derived and the position of the
zero-electric-field resonance is determined at
$35,648.5(1)$~cm$^{-1}$. As shown in Fig.~\ref{fig:SmallRangeEd},
no state has been previously identified with this energy. The
variation of the amplitude of the signal as a function of the
polarizations of the laser beams is consistent with that of a
$M=0$ sublevel. This ``new" odd-parity state is probably the
missing 6s8p~$^3$P$_0$ state. The existence of this state explains
the large polarizabilities of the 6s7d~$^3$D$_1$ state.

\begin{figure}
\includegraphics[width=3.25 in]{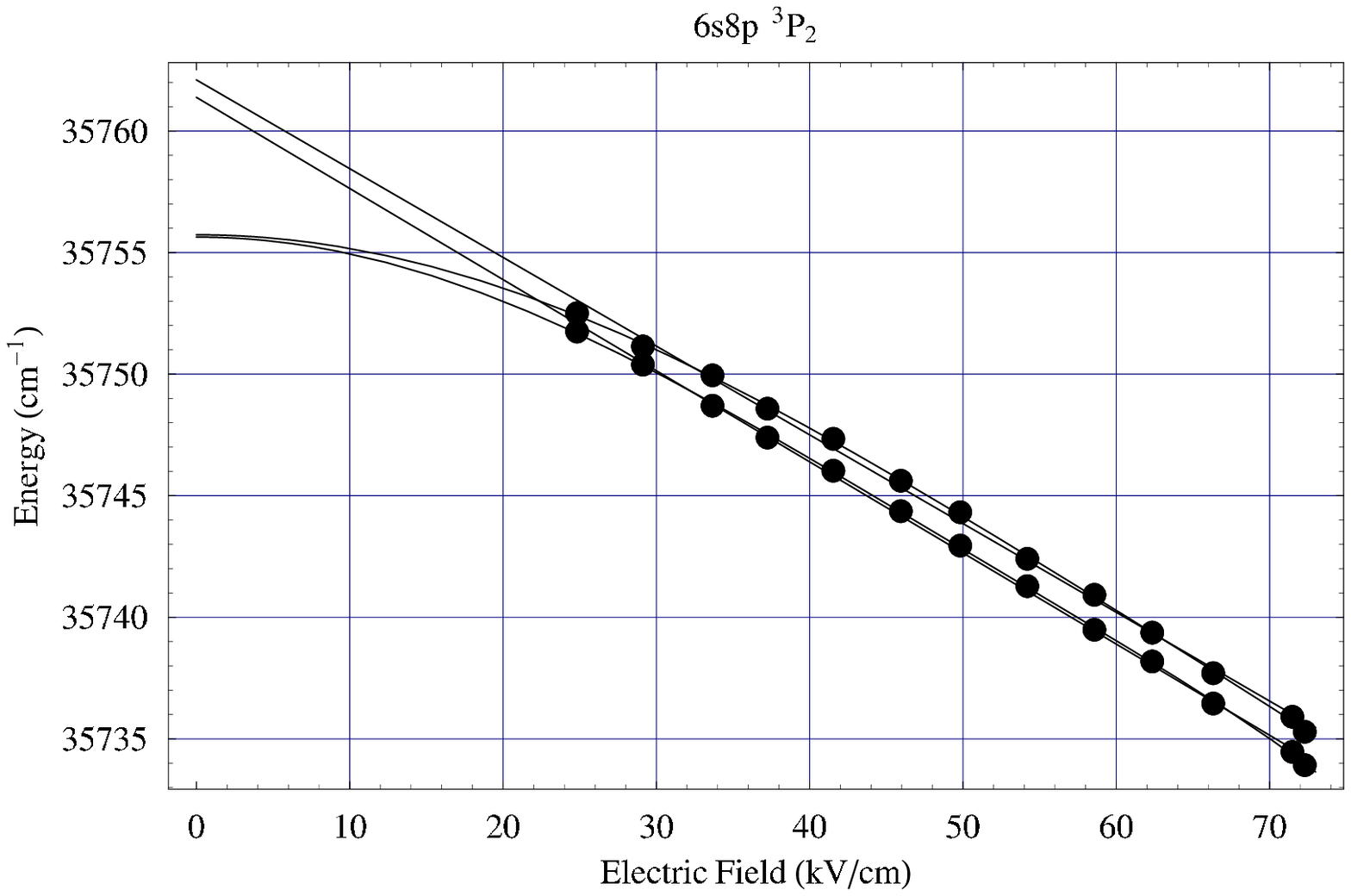}
\caption{The Stark splitting and shift for the resonance of the
6s8p~$^3$P$_2$ state. The data points corresponding to $|M|=1$ and
$|M|=2$ sublevels are fit by both a linear function and a
sixth-order polynomial functions. The linear fit gives an upper
bound of the energy of 6s8p~$^3$P$_2$ state while the sixth-order
polynomial fit gives a lower bound.} \label{fig:Ba6s8p3P2}
\end{figure}

The other two of the Stark-induced resonances are classified as
the third set. We have identified them (using the
laser-polarization dependence of the signal amplitude) as due to
the $|M|=1$ and $|M|=2$ sublevels, respectively. From the
electric-field dependence of the amplitudes of the signals, the
perturbation theory is not applicable in either case. This is
consistent with that the Stark-induced energy shifts are almost
linearly dependent on the applied electric field
(Fig.~\ref{fig:Ba6s8p3P2}). The data show that the
zero-field-energy of a $J\ge2$ odd-parity state should be within
$35,756$~cm$^{-1}$, derived from polynomial fit, and
$35,761$~cm$^{-1}$, derived from linear fit. This is consistent
with the predicted energy of the 6s8p~$^3$P$_2$ state
($35,756(1)$~cm$^{-1}$) from the Stark-effect of the
6s7d~$^3$D$_2\ |M|=2$ sublevels with the three-level
approximation. Therefore, these two Stark-induced resonances are
probably due to the $|M|=1$ and $|M|=2$ sublevels of the
$6$s$8$p~$^3$P$_2$ state. With a ten-level model, which includes
all the states with the energies between $35,600$ and
$36,000$~cm$^{-1}$, the energy of the 6s8p~$^3$P$_2$ state is
determined at $35,757(1)$~cm$^{-1}$. The $M=0$ sublevel is missing
because, unlike the $|M|=1$ and $|M|=2$ sublevels, it cannot
couple to the most closely lying state, 6s7d~$^3$D$_2$. Therefore,
the Stark-induced transition is weaker. While the fluorescence
resulting from the Stark-induced transitions to the $|M|=1$ and
$|M|=2$ sublevels is already feeble, it is too weak to be
detected. The polarizabilities of the 6s8p~$^3$P$_2$ state is
difficult to determine because of the lack of the low-field data.
However, the lower bounds of the scalar and tensor
polarizabilities can be determined
(Table~\ref{table:polarizability}).

\section{Data analysis}

In subsection A, we estimate the mixing of states according to the
known energy spectrum, including the newly found states. In
subsection B, we derive the electric dipole moments from the
measured polarizabilities in Table~\ref{table:polarizability}.

\subsection{Configuration and spin-orbit mixing}

There are two multiplets, 6s7d~$^3$D$_J$ and 6s8p~$^3$P$_J$, in
the energy interval under discussion. Both present some
irregularities in the fine-structure splittings. We can use them
to estimate the mixing with the nearby levels.

The ratio of the ``unperturbed" intervals for the 6s7d~$^3$D$_J$
multiplet should be $(E_{J=3}-E_{J=2})/(E_{J=2}-E_{J=1}) = 3/2$,
according to the Lande interval rule \cite{Sobelman}. The most
probable explanation for the deviation from this rule is the
repulsion between the 6s7d~$^3$D$_2$ and 6p$^2\
^3$P$_2$\footnote{The interaction with the 6s7d~$^1$D$_2$ state is
suppressed because of the relatively large energy difference.}.
Assuming that 6p$^2\ ^3$P$_2$ is the only perturbing source of
this multiplet, we have a simplified Hamiltonian for these two
states:
\begin{equation}
\left( \begin{array}{cc}
E_0(6\rm{p}^2\ ^3\rm{P}_2) & \mathcal{V}\\
\mathcal{V}  & E_0(6\rm{s}7\rm{d}\ ^3\rm{D}_2)
\end{array} \right),
\label{Eqn:mixingsecular}
\end{equation}
where the $E_0$'s are the ``unperturbed" energies of 6p$^2\
^3$P$_2$ and 6s7d~$^3$D$_2$ states and $\mathcal{V}= \langle
6$s7d~$^3$D$_2|V|6$p$^2\ ^3$P$_2\rangle$, where $V$ is the
potential that causes the configuration and spin-orbit mixing. The
eigenstates of the matrix are the mixtures of these two states. If
we write them in a column vector, they can be expressed as:
\begin{equation}
\left( \begin{array}{c}
\rm ``6 p^2\ ^3P_2"\\
\rm ``6s7d\ ^3D_2"
\end{array} \right)=
\left( \begin{array}{cc}
{\rm cos} \phi & {\rm sin} \phi\\
-{\rm sin} \phi  & {\rm cos} \phi
\end{array} \right)
\left( \begin{array}{c}
\rm 6p^2\ ^3P_2\\
\rm 6s7d\ ^3D_2
\end{array} \right),
\label{Eqn:mixingsecular}
\end{equation}
where $\phi={\rm
arctan}\{2\mathcal{V}/[E_0(6$s7d~$^3$D$_2)-E_0(6$p$^2\
^3$P$_2)]\}/2$ is the mixing angle. The mixing ``shifts" the
energies of the states by
[sec$(2\phi)-1]\cdot[E_0(6$s7d~$^3$D$_2)-E_0(6$p$^2\ ^3$P$_2)]/2$
in ``repulsive" directions. Assuming that $6$s$7$d~$^3$D$_{1,3}$
states are unperturbed, we estimate the shift of the level
6s7d~$^3$D$_2$ to be $23$~cm$^{-1}$. That gives the mixing angle
$\phi \approx {\rm arctan}(0.44)$. This two-level model can
plausibly explain the large electric dipole moments between the
6p$^2\ ^3$P$_2$ state and the 6s8p~$^3$P$_{1,2}$ states since it
predicts:
\begin{equation}\label{c2}
  \frac{\langle \rm 6s7d\ ^3D_2|{\it D}|6s8p\ ^3P_1\rangle}
  {\langle \rm 6p^2\ ^3P_2|{\it D}|6s8p\ ^3P_1\rangle}
  = \frac{\langle \rm 6s7d\ ^3D_2|{\it D}|6s8p\ ^3P_2\rangle}
  {\langle \rm 6p^2\ ^3P_2|{\it D}|6s8p\ ^3P_2\rangle}
  = {\rm tan}\phi.
\end{equation}

Similar analysis shows that there is an interaction between
6s8p~$^3$P$_1$ and 6s8p~$^1$P$_1$ states. In the $LS-$coupling
scheme, assuming the coupling with the rest of the states is
negligible, we have a simplified Hamiltonian for the
6s8p~$^3$P$_{0,1,2}$ and $^1$P$_1$ states:\cite{Condon}
\begin{equation}
\begin{array}{c}
\begin{array}{cccccc}
&~~ & ^3{\rm P}_2~~~~~~~~ & ^3{\rm P}_1~~~~~~~~ & ^1{\rm
P}_1~~~~~~~~ & ^3{\rm P}_0~
\end{array}\\
\begin{array}{c}
^3{\rm P}_2\\
^3{\rm P}_1\\
^1{\rm P}_1\\
^3{\rm P}_0
\end{array}
\left( \begin{array}{cccc}
F-G-\zeta & 0 & 0 & 0\\
0  & F-G-\zeta & \sqrt{2}~\zeta &0\\
0 & \sqrt{2}~\zeta & F+G & 0\\
0 & 0 & 0 & F-G-2~\zeta\\
\end{array} \right),
\end{array}
\label{Eqn:mixingsecular}
\end{equation}
where $F$ is an additive constant, $G$ causes the splitting of the
triplet and the singlet, and $\zeta$ is from the spin-orbit
interaction that causes the splitting of the $^3$P$_J$ states and
the mixing of the $^3$P$_1$ and the $^1$P$_1$ states. There are
only three parameters to determine four energies of the 6s8p
multiplets. The eigenvalues of this Hamiltonian fit the observed
energies of the 6s8p multiplets very well, deviated by
$<2$~cm$^{-1}$ compared to the $\sim 100$~cm$^{-1}$ energy
difference between these four states. The consistency of the
$LS-$coupling scheme with the observed energies of the 6s8p
multiplets indicates that the configuration mixing is
small.\footnote{Similar analysis of 6s6p multiplets in Ba (and
also in other atoms such as Yb) shows larger deviations from the
experimental data, which can be explained by stronger
configuration mixing due to larger overlap of the wavefunctions of
the 6s, 6p, and 5d electrons.} The mixing angle between the
$^3$P$_1$ and the $^1$P$_1$ states is $\chi \approx {\rm
arctan}(0.27)$.

For both cases, because of the non-negligible mixing angles, the
mixing should be taken into account in the analysis in the next
subsection.

In the $LS$-coupling scheme, the ratios between the electric
dipole moments of the 6s7d~$^3$D$_J$ and 6s8p~$^3$P$_J$ multiplets
are given in Table~\ref{multiplets}~\cite{Sobelman}. Since $LS$
coupling is not exact, we do not expect these relations to be
accurate. However, they can give us the estimates for the
contributions of the weak couplings.

\subsection{Electric dipole moments derived from experimental data on polarizabilities}

In this subsection, we try to derive electric dipole moments from
the measured polarizabilities presented in
Table~\ref{table:polarizability}. In Table~\ref{tab1}, the angular
coefficients from Eqs.(\ref{a3},\ref{a4}) are tabulated. It
simplifies the analysis if one uses combinations of $\alpha_0$ and
$\alpha_2$ in order to exclude contributions of particular
opposite-parity states. These combinations for $J=1$ and $J=2$ are
given in Table~\ref{tab2}. The dipole moments derived in this
subsection are listed in Table~\ref{table:deriveE1}.

\subsubsection{The $\rm 6p^2\ ^3P_2$ state.} The three dominant coupling
states are 6s8p~$^3$P$_{1,2}$, and 6s8p~$^1$P$_1$. The
5d7p~$^3$F$_2$ state is negligible because energy difference is
large and the configuration is different by two electrons with
$\Delta L=2$. Using Table~\ref{tab2}, we can exclude the
contribution from $J=1$ states by taking the following combination
of the scalar and tensor polarizabilities:
\begin{equation}\label{b1}
\alpha_0 + \alpha_2 = +\frac{4}{15} \frac{|\langle 6\rm{p^2\
^3P}_2||\it{D}||\rm{6s8p\ ^3P}_2\rangle|^2} {140(1)~{\rm
cm}^{-1}}.
\end{equation}
This gives the dipole moment $|\langle 6$p$^2\
^3$P$_2||D||$6s8p~$^3$P$_2\rangle| = 6.0(2)$~ea$_0$. Similarly,
the contribution from $J=2$ states can be excluded in a different
combination of the polarizabilities.
\begin{equation}\label{b3}
\alpha_0 - \alpha_2 = \frac{4}{15}\left( \frac{|\langle \rm 6p^2\
^3P_2||{\it D} ||6s8p\ ^3P_1\rangle|^2} {52.0(2)~{\rm cm}^{-1}}+
\frac{|\langle \rm 6p^2\ ^3P_2||{\it D}||6s8p\ ^1P_1\rangle|^2}
{276~{\rm cm}^{-1}}\right).
\end{equation}
The contribution of the 6s8p~$^1$P$_1$ state can be estimated:
\begin{eqnarray}\label{b4a}
&&|\langle \rm 6p^2\ ^3P_2||{\it D}||6s8p\ ^1P_1\rangle|^2 \approx
\tan^2 \chi\cdot |\langle 6p^2\ ^3P_2||{\it D}||6s8p\
^3P_1\rangle|^2,
\end{eqnarray}
This gives the dipole moment $|\langle 6$p$^2\
^3$P$_2||D||6$s8p~$^3$P$_1\rangle| = 14.34(6)$~ea$_0$. The
estimate, Eq.(\ref{b4a}), leads to a $0.7\%$ decrease of the
dipole moment. Therefore, it is unlikely that this approximation
leads to a large error.

\subsubsection{The $\rm 6s7d\ ^3D_1$ state.} The four dominant coupling
states are the 6s8p~$^3$P$_{0,1,2}$, and 6s8p~$^1$P$_1$ states.
The coupling to the 5d7p~$^3$F$_2$ state is negligible because the
configurations is different by two electrons and the energy
difference is large. The coupling to the 6s8p~$^3$P$_2$ state is
suppressed because $\Delta J=-\Delta L$, which is strongly
suppressed (Table~\ref{multiplets}). By taking the following
combination:
\begin{eqnarray}
\label{b5a} 2\alpha_2 - \alpha_0 &=& +\frac{2}{3} \frac{|\langle
{\rm 6s7d\ ^3D_1}|| D ||{\rm6s8p\ ^3P_0}\rangle|^2} {60.8(1)~{\rm
cm}^{-1}} -\frac{4}{15} \frac{|\langle {\rm 6s7d\
^3D_1}||D||{\rm6s8p\ ^3P_2}\rangle|^2} {48(1)~{\rm cm}^{-1}}
\\ \label{b5b}
&\approx& (1-0.025) \frac{|\langle {\rm 6s7d\ ^3D_1}||D||{\rm
6s8p\ ^3P_0}\rangle|^2} {91.2(2)~{\rm cm}^{-1}},
\end{eqnarray}
we can derive $|\langle 6$s7d~$^3$D$_1||D||6$s8p~$^3$P$_0\rangle|
= 15.9(1)$~ea$_0$. The estimated contribution of the
6s8p~$^3$P$_2$ state changes the dipole moment by $\sim 1\%$,
which justifies the estimate.

Another combination of the polarizabilities gives the second
equation for this state:
\begin{eqnarray}
\label{b6a} \alpha_0 + \alpha_2 &=& -\frac{1}{3}\frac{\rm |\langle
6s7d\ ^3D_1||{\it D}||6s8p\ ^3P_1\rangle|^2} {\rm 40.3(2)~cm^{-1}}
+ \frac{1}{5}\frac{\rm |\langle 6s7d\ ^3D_1||{\it D}||6s8p\
^3P_2\rangle|^2}
{\rm 48(1)~cm^{-1}}\\
&&+ \frac{1}{3}\frac{\rm |\langle 6s7d\ ^3D_1||{\it D}||6s8p\
^1P_1\rangle|^2} {\rm 183~cm^{-1}}
\end{eqnarray}
Again, we estimate $|\langle
6$s7d~$^3$D$_1||D||6$s8p~$^3$P$_2\rangle|$ with
Table~\ref{multiplets} and $|\langle
6$s7d~$^3$D$_1||D||6$s8p~$^1$P$_1\rangle|^2 \approx {\rm
tan}^2\chi \cdot |\langle
6$s7d~$^3$D$_1||D||6$s8p~$^3$P$_1\rangle|^2$. This derives the
dipole moment $|\langle 6$s7d~$^3$D$_1||D||6$s8p~$^3$P$_1\rangle|
= 12.4(2)~$ea$_0$. Note that the ratio of the dipole moments
$|\langle 6$s7d~$^3$D$_1||D||6$s8p~$^3$P$_0\rangle|$ and $|\langle
6$s7d~$^3$D$_1||D||6$s8p~$^3$P$_1\rangle|$ is 1.28(1), which
differs only by 10$\%$ from the $LS$-coupling prediction
(Table~\ref{tab_ls}).

\subsubsection{The $\rm 6s7d\ ^3D_2$ state.} The possible dominant coupling
states are 6s8p~$^3$P$_{1,2},\ ^1P_1$ and 5d7p~$^3$F$_2$ states.
Again, we can take the different combinations of the scalar and
tensor polarizabilities to extract the contribution of $J=1$ or
$J=2$ states.
\begin{equation}
\label{b7a} \alpha_0 + \alpha_2 = -\frac{4}{15}\left(
\frac{|\langle \rm 6s7d\ ^3D_2||{\it D}||6s8p\ ^3P_2\rangle|^2}
{5(1)~{\rm cm}^{-1}} + \frac{|\langle \rm 6s7d\ ^3D_2||{\it
D}||5d7p\ ^3F_2\rangle|^2} {-473~{\rm cm}^{-1}}\right),
\end{equation}
We can neglect the second term in the right hand side of
Eq.(\ref{b7a}) considering the relatively large uncertainty in the
denominator of the first term. It follows that $|\langle
6$s7d~$^3$D$_2||D||6$s8p$^3$P$_2\rangle| =14(2)$~ea$_0$. Note that
Table~\ref{tab_ls} predicts $|\langle
6$s7d~$^3$D$_2||D||6$s8p~$^3$P$_2\rangle|=|\langle
6$s7d~$^3$D$_1||D||6$s8p~$^3$P$_1\rangle|$, so we have reasonable
agreement with the $LS$-coupling. The second equation is:
\begin{eqnarray}
\label{b8a} \alpha_2 - \alpha_0 &=&\frac{4}{15}\left( \frac{\rm
|\langle 6s7d\ ^3D_2||{\it D}||6s8p\ ^3P_1\rangle|^2}
{93.2(2)~{\rm cm}^{-1}} + \frac{|\langle \rm 6s7d\ ^3D_2||{\it
D}||6s8p\ ^1P_1\rangle|^2} {-130~{\rm cm}^{-1}}\right)
\\
\label{b8b} &\approx&(1-0.052) \frac{|\langle \rm 6s7d\
^3D_2||{\it D}||6s8p\ ^3P_1\rangle|^2} {348.8~{\rm cm}^{-1}}.
\end{eqnarray}
This gives the dipole moment $|\langle
6$s7d~$^3$D$_2||D||6$s8p~$^3$P$_1\rangle| =  21(3)$~ea$_0$. The
ratio between $|\langle 6$s7d~$^3$D$_2||D||6$s8p~$^3P_1\rangle|$
and $|\langle 6$s7d~$^3$D$_2||D||6$s8p~$^3$P$_2\rangle|$ is
1.5(4), which is consistent with the prediction of
Table~\ref{multiplets}. The ratio between $|\langle 6$p$^2\
^3$P$_2||D||6$s8p~$^3$P$_1\rangle|$ and $|\langle
6$s7d~$^3$D$_2||D||6$s8p~$^3$P$_1\rangle|$ is 0.7(1), which is
somewhat larger than the prediction of Eq.~(\ref{c2}), $\rm tan
\phi \approx 0.44$.

\subsubsection{The $\rm 5d6d\ ^3D_1$ state.}
For this state, we can expect that the largest contribution of the
polarizabilities is from the coupling to the 5d7p~$^3$F$_2$ state,
which is counterbalanced by four smaller couplings to the closer
levels. This may explain the smallness of both polarizabilities of
this level. However, the quantitative analysis is hampered by the
large number of contributions. Thus, we are not able to use
experimental data for this state in our analysis.

\subsubsection{The $\rm 6s8p\ ^3P_0$ state.}
The dominant coupling states are the 6s7d~$^3$D$_1$ and
5d6d~$^3$D$_1$ states. We can use Eq.(\ref{b5b}) to calculate the
contribution of the 6s7d~$^3$D$_1$ state and estimate the coupling
to the 5d6d~$^3$D$_1$ state.
\begin{eqnarray}
\label{b9a} \alpha_0 &=& \prn{-\frac{2}{3}} \prn{\frac{|\langle
\rm 6s7d\ ^3D_1||{\it D}||6s8p\ ^3P_0\rangle|^2} {-60.8(1)~{\rm
cm}^{-1}}+\frac{|\langle \rm 5d6d\ ^3D_1||{\it D}||6s8p\
^3P_0\rangle|^2} {-285.3(1)~{\rm cm}^{-1}}}
\\
\label{b9b}  & \approx & \frac{1}{0.975} \left(2\alpha_2 -
\alpha_0\right)_{\rm 6s7d\ ^3D_1} + \frac{2}{3} \frac{|\langle \rm
5d6d\ ^3D_1||{\it D}||6s8p\ ^3P_0\rangle|^2} {285.3(1)~{\rm
cm}^{-1}}
\end{eqnarray}
This gives the dipole moment $|\langle
5$d6d~$^3$D$_1||D||6$s8p~$^3$P$_0\rangle| = 14(2)$~ea$_0$, which
is unexpectedly large considering the configurations of these two
states are different by two electrons. This large dipole moment
probably means that the 5d6d~$^3$D$_1$ state interacts with the
6s7d~$^3$D$_1$ state. This interaction can influence the mixing
angle $\phi$, which is derived under the assumption that the
6s7d~$^3$D$_{0,2}$ states are unperturbed by the configuration or
spin-orbit mixing. This interaction ``shifts" the 6s7d~$^3$D$_2$
state downwards and makes mixing angle $\phi$ larger, which is
consistent with our expectation of larger $\rm tan(\phi)$ to
explain the large $|\langle 6$p$^2\
^3$P$_2||D||6$s8p~$^3$P$_1\rangle|$ dipole moment.

\subsubsection{The $\rm 6s8p\ ^3P_1$ state.} The dominant coupling states
for the 6s8p~$^3$P$_1$ state are 6p$^2\ ^3$P$_2$, 6s7d~$^3$D$_1$,
and 5d6d~$^3$D$_1$ states. We have derived all major contributions
to the polarizabilities of the 6s8p~$^3$P$_1$ state. Therefore, we
can use the polarizabilities to check our model for consistency.
The scalar polarizability is given by the following expression:
\begin{eqnarray}
\label{b10} \alpha_0 & = & -\frac{2}{9}\left( \frac{|\langle \rm
6p^2\ ^3P_2||{\it D}||6s8p\ ^3P_1\rangle|^2} {52.0(2)~{\rm
cm}^{-1}} +\frac{|\langle \rm 6s7d\ ^3D_1||{\it D}||6s8p\
^3P_1\rangle|^2} {-40.3(2)~{\rm cm}^{-1}} \right.\nonumber
\\
&&+\left. \frac{|\langle \rm 6s7d\ ^3D_2(||{\it D}||6s8p\
^3P_1\rangle|^2} {-93.2(2)~{\rm cm}^{-1}} +\frac{|\langle \rm
5d6d\ ^3D_1||{\it D}||6s8p\ ^3P_1\rangle|^2} {-265~{\rm cm}^{-1}}
\right).
\end{eqnarray}
Using the experimental data in Table~\ref{table:polarizability}
and Eqs.(\ref{b3},\ref{b6a},\ref{b8b}), we have:
\begin{equation}
\frac{|\langle \rm 5d6d\ ^3D_1||{\it D}||6s8p\ ^3P_1\rangle|^2}
{1193~{\rm cm}^{-1}} = 0.8(5)~{\rm ea_0^2/cm^{-1}}. \label{b11}
\end{equation}
This is almost consistent with zero contribution from the
5d6d~$^3$D$_1$ state. However, because of the large error in
Eq.(\ref{b11}), we cannot obtain significant bound on the dipole
moment $|\langle 5$d6d~$^3$D$_1||D||6$s8p~$^3$P$_1\rangle|$. The
tensor polarizability can be expressed as:
\begin{eqnarray}
\label{b13} \alpha_2 & = & \frac{1}{45} \frac{|\langle \rm 6p^2\
^3P_2||{\it D}||6s8p\ ^3P_1\rangle|^2} {52~{\rm cm}^{-1}}
-\frac{1}{9} \frac{|\langle \rm 6s7d\ ^3D_1||{\it D}||6s8p\
^3P_1\rangle|^2} {-40~{\rm cm}^{-1}}
\nonumber\\
&&+\frac{1}{45} \frac{|\langle \rm 6s7d\ ^3D_2||{\it D}||6s8p\
^3P_1\rangle|^2} {-93~{\rm cm}^{-1}} -\frac{1}{9} \frac{|\langle
\rm 5d6d\ ^3D_1||{\it D}||6s8p\ ^3P_1\rangle|^2} {-265~{\rm
cm}^{-1}}.
\end{eqnarray}
With similar method, we get:
\begin{equation}
\frac{|\langle \rm 5d6d\ ^3D_1||{\it D}||6s8p\ ^3P_1\rangle|^2}
{2385~{\rm cm}^{-1}} \approx 0.0(1)~{\rm ea_0^2/cm^{-1}}.
\end{equation}
This is consistent with the zero contribution from the
5d6d~$^3$D$_1$ state and gives an upper bound:
\begin{equation}
|\langle \rm 5d6d\ ^3D_1||{\it D}||6s8p\ ^3P_1\rangle| \le
15~ea_0.
\end{equation}

\section{Systematic uncertainties}

The systematic uncertainties in this work come in two ways. One is
from the simplified models used in the data analysis. The other is
from the limited sensitivities of the apparatus. It is found that
the systematic uncertainties are mostly from the simplified
models. To take them into account, we multiply the statistical
uncertainties by the square-root of the reduced~$\chi^2$ to be the
presented uncertainties.

The possible sources of the systematic uncertainties from the
limited sensitivity of the apparatus are discussed. It is found
that the dominant errors come from the drift of the frequency
markers (in the first part of the experiment) and the uncertainty
in the reading of the grating position (in the second part of the
experiment).

In the first part of the experiment, a temporal drift of the
frequency-markers was observed. The overall drift is about $10\
$GHz over $\sim 4$~hours of run time. The effect is consistent
with what we expect from the thermal variation of the index of
refraction and the thermal expansion of the glass heated by the
laser and/or variations of room temperature. To account for the
drift, we recorded zero-electric-field scans before and after each
non-zero-field scan. The drifts of the frequency markers relative
to the positions of the zero-electric-field resonance peaks were
thus observed and recorded. Because the drift is a smooth function
of time, knowing at what time a given scan is recorded, we can
determine the reference point of the energy shifts when the
electric field is applied from the positions of the two adjacent
zero-electric-field resonance peaks. This systematic error can be
reduced to $\approx 500$~MHz. However, this is still the dominant
source of the systematic uncertainties.

In the second part of the experiment, it is more difficult to
determine the Stark-induced energy shift because the signal is
only present when the electric field is applied. In order to
reduce the error from the temporal drift of the reading of the
grating position, we use the same method, but instead of recording
the position of a resonance at zero electric field, we record that
of transition to 6p$^2\ ^3$P$_2$, whose energy is well-determined,
before and after each measurement of the resonance of the
Stark-induced transition. This method reduces the systematic error
to $\approx 6$~GHz. This is the dominant source of the systematic
uncertainties.

The electric field calibration was discussed in Ref.\cite{Roc99}.
The high voltage output is calibrated to $0.015\%$. The electrode
spacing is calibrated to $0.02\%$. This determines the electric
field with $0.03\%$ precision. Therefore, the systematic
uncertainty of the electric field is negligible.

In the presence of strong laser pulses, the energy of a state is
shifted, due to the dynamic Stark effect. To avoid this effect, we
temporally separate the two pulses by $20-30$~ns and attenuate the
powers of lasers, so they barely saturate the transitions. We have
verified that the measured polarizabilities are independent of the
temporal separation of the two laser pulses or the laser powers,
so the dynamic Stark effect is negligible in our experiments.

The errors from the electronic noises, the uncertainties of the
readouts of the barometer and CAMAC modules, and the hyperfine
structures of the barium isotopes with non-zero nuclear spin
($^{135}$Ba and $^{137}$Ba, both with I$=3/2$) are also estimated.
It is found that these systematic errors are negligible.

\section{Conclusions}
\label{Section:Conclus}

In this work, we have measured the scalar and tensor
polarizabilities of four even-parity states of barium ranging from
35,600 to 36,000~cm$^{-1}$, and the hyperpolarizabilities of two
of those four states. Three of these states have unusually large
polarizabilities which exceed the value that might be expected
from the known energy levels of barium by more than two orders of
magnitude. Our experimental data suggest the existence of two
``new" states. These two states have been identified by direct
laser spectroscopy of the Stark-induced transitions and their
energies have been determined to within $\approx 1$~cm$^{-1}$. The
polarizabilities of two odd-parity states excited via the
Stark-induced transitions are measured. Using the polarizabilities
measured in this work, we have also derived seven electric dipole
moments. They relative values are consistent with the prediction
of the $LS-$coupling scheme.

\acknowledgments

The authors wish to thank V. V. Yashchuk and D. English for help
with the experiments and useful discussions, and D. P. DeMille, B.
P. Das, W. Gawlik, W. C. Martin and M. Kuchiev for helpful advice.
This research was supported by the National Science Foundation
Career grant PHY-9733479 and the Miller Institute for Basic
Research in Science.

\bigskip

\begin{table} \caption{The relation between the polarizations of laser
beams and the excited Zeeman sublevels of a state with total
angular momentum $J$, which are excited by two E1 transitions from
the $J=0$ ground state via a $J=1$ intermediate state.}
\medskip \begin{tabular}{|c||c|c|c|} \hline
\hline & \multicolumn{3}{|c|}{Polarizations of laser beams}\\
\cline{2-4} $J$ & z, z & y, z or z, y & y, y     \\
\hline
\hline      $0$ & $M=0$ &    -     & $M=0$   \\
\hline      $1$ & -     & $|M|=1$  &  -      \\
\hline      $2$ & $M=0$ & $|M|=1$  & $|M|=0,2$ \\
\hline \hline
\end{tabular}
\label{table:laserpol}
\end{table}

\begin{table}
\caption{Observed scalar and tensor polarizabilities, in units of
MHz/(kV/cm)$^2$}
\medskip \begin{tabular}{|c|c|c|} \hline \hline
State & $\alpha_0$& $\alpha_2$\\
\hline 6p$^2\ ^3$P$_2$ & $31.05(9)$ & $-27.3(1)$\\
\hline 6s7d~$^3$D$_1$ & $-93(1)$   & $26.6(5)$ \\
\hline 6s7d~$^3$D$_2$ & $-335(13)$ & $-268(13)$ \\
\hline 5d6d~$^3$D$_1$ & $-4.2(2)$  & $2.0(1)$ \\
\hline 6s8p~$^3$P$_0$ & $176(3)$   & $-$ \\
\hline 6s8p~$^3$P$_1$ & $97(6)$    & $24(3)$ \\
\hline 6s8p~$^3$P$_2$ & $\ge 370$  & $\ge 60$ \\
\hline \hline
\end{tabular}
\label{table:polarizability}
\end{table}

\begin{table}
\caption{Observed scalar and tensor hyperpolarizabilities in units
of kHz/(kV/cm)$^4$}
\medskip \begin{tabular}{|c|c|c|c|} \hline \hline
        State         & $\gamma_0$ & $\gamma_2$   &  $\gamma_4$\\
\hline  6p$^2\ ^3$P$_2$ & $-2.2(16)$      & $0(1)$      &    $-0.6(3)$   \\
\hline  6s7d~$^3$D$_1$ & $61(3)$        & $-5(2)$          &    $-$                 \\
\hline \hline
\end{tabular}
\label{table:hyperpolarizability}
\end{table}

\begin{table} \caption{Estimated reduced electric dipole moments in
units of ea$_0$ with the two- or three-level approximation.}
\medskip \begin{tabular}{|c|c|c|} \hline
\hline Reduced dipole moment & Experiment & Bates-Damgaard \\
\hline $\rm |(6p^2\ ^3P_2\|{\it D}\|6s8p\ ^3P_1)|_{\rm two-level}$\footnotemark[1]  & $14.5(3)$ & N/A\\
\hline $\rm |(6s7d\ ^3D_2\|{\it D}\|6s8p\ ^3P_1)|_{\rm two-level}$   & $22(2)$  & $34.7$\\
\hline $\rm |(6s7d\ ^3D_2\|{\it D}\|6s8p\ ^3P_2)|_{\rm three-level}$\footnotemark[1] & $16.2(5)$ & $15.5$\\
\hline $\rm |(6s7d\ ^3D_3\|{\it D}\|6s8p\ ^3P_2)|_{\rm three-level}$\footnotemark[1] & $40(2)$   & $44$\\
\hline \hline
\end{tabular}
\label{table:dipole} \footnotetext[1]{The levels are designated by
their nominal or proposed configurations (see text).}
\end{table}

\begin{table}
\caption{Relative values of the reduced matrix elements between
the levels of the 6s7d~$^3$D$_J$ and 6s8p~$^3$P$_J$ multiples in
pure $LS$-coupling.} \label{tab_ls}
\begin{tabular}{|c|c|c|c|}
\hline \hline &\multicolumn{1}{c|}{6s8p~$^3$P$_0$}
&\multicolumn{1}{c|}{6s8p~$^3$P$_1$}
&\multicolumn{1}{c|}{6s8p~$^3$P$_2$}\\
\hline
\hline 6s7d~$^3$D$_1$  & $\sqrt{20}$ & $\sqrt{15}$ &     1           \\
\hline 6s7d~$^3$D$_2$  &     0       & $\sqrt{45}$ & $\sqrt{15}$     \\
\hline 6s7d~$^3$D$_3$  &     0       &      0      & $\sqrt{84}$     \\
\hline \hline
\end{tabular}
\label{multiplets}
\end{table}

\begin{table}[ht]
\caption{Angular coefficients for $\alpha_0$ and $\alpha_2$.}
\label{tab1}
\begin{tabular}{|c|cc|cc|cc|cc|}
\hline \hline $J_0~:~J_1$&\multicolumn{2}{c|}{0}
&\multicolumn{2}{c|}{1} &\multicolumn{2}{c|}{2}
&\multicolumn{2}{c|}{3} \\
\hline
\hline 0 &$-$2/3 & 0   &$-$2/3 & 0     &       &       &       &        \\
\hline 1 &$-$2/9 & +2/9&$-$2/9 &$-$1/9 &$-$2/9 & +1/45 &       &        \\
\hline 2 &       &     &$-$2/15& +2/15 &$-$2/15&$-$2/15&$-$2/15& +4/105 \\
\hline 3 &       &     &       &       &$-$2/21& +2/21 &$-$2/21&$-$5/42 \\
\hline \hline
\end{tabular}
\end{table}

\begin{table}[htb]
\caption{Combinations of $\alpha_0$ and $\alpha_2$ that can
simplify the analysis.}

\label{tab2}

\begin{tabular}{|c|c|c|}
\hline \hline
     $J_0=1$            &     $J_0=2$            & does not depend on \\
\hline
\hline $\alpha_0 + \alpha_2$ &                      & $J_1=0$, \\
\hline $\alpha_0 -2\alpha_2$ &$\alpha_0 + \alpha_2$ & $J_1=1$, \\
\hline $\alpha_0 +10\alpha_2$&$\alpha_0 - \alpha_2$ & $J_1=2$, \\
\hline                      &$2\alpha_0 +7\alpha_2$ & $J_1=3$. \\
\hline \hline
\end{tabular}
\end{table}

\begin{table}
\caption{Electric dipole moments (in units of ea$_0$) derived from
the polarizabilities measured in this work.}
\medskip\begin{tabular}{|c|c|c|c|} \hline
\hline & 6s8p~$^3$P$_0$ & 6s8p~$^3$P$_1$ & 6s8p~$^3$P$_2$\\
\hline
\hline 6p$^2\ ^3$P$_2$ & 0        & 14.34(6) & 6.0(2)\\
\hline 6s7d~$^3$D$_1$ & 15.9(1)  & 12.4(2)  & -\\
\hline 6s7d~$^3$D$_2$ & 0        & 21(3)    & 14(2)\\
\hline 5d6d~$^3$D$_1$ & 14(2)    & $\le$ 15 &  - \\
\hline\hline
\end{tabular}
\label{table:deriveE1}
\end{table}

\bibliography{Ba}
\end{document}